# Survey on Using GIS in Evacuation Planning Process

Sara Shaker Abed El-Hamied

Information Systems Department
Computer and Information Sciences
Faculty
Mansoura, Egypt
sara.shaker2008@yahoo.com

Ahmed Abou El-Fotouh Saleh

Information Systems Department
Computer and Information Sciences
Faculty
Mansoura, Egypt
elfetouh@gmail.com

Aziza Asem

Information Systems Department
Computer and Information Sciences
Faculty
Mansoura, Egypt
aziza_asem2@hotmail.com

*Abstract-* **Natural crises form a big threat on environment; these crises mean the loss of enterprises and individuals, and therefore losses in the sum total of community development. Management to these crises is required through a crisis management plan to control the crises before, during, and after the event. One of the most needed things to consider during preparing the crisis management plan is preparing the evacuation plan in order to transfer people from the incident place to a safe place; this must be done quickly and carefully.**
**Because of the geographic nature of the evacuation process, Geographical Information System (GIS) has been used widely and effectively for over 20 years in the field of crisis management in general and in evacuation planning in particular. This paper provides an overview about evacuation process and the basic concepts of GIS systems. The paper also demonstrates the importance of evacuation planning and how GIS systems used in other studies to assists in evacuation process**

## I. INTRODUCTION

Crises always threaten society; they happen suddenly and cause significant losses. Crisis management is the process of controlling crises before, during, and after the event. One of crisis management activities is Evacuation planning, which means the transfer of people from an unsafe place to another safe place.

Within crises management, emergency management applies geo-information technologies in the crisis management process and Geographical Information Systems (GIS) have been used for over 20 years. Examples for GIS utilization in natural and man-made disasters are to support flood mapping, hurricane prediction, and environmental clean-ups after industrial accidents.

In most crises situations, GIS operators receive their orders via staff members who are asked by the decision makers to inquire about maps. The GIS specialists usually react to mapping and spatial analysis requests from decision makers, e.g. after the World Trade Center attack GIS specialists, supported by company consultants, were operating Geographic Information Systems and producing maps on demand and, after

Hurricane Katrina, GIS experts from Louisiana State University provided support to evacuation and relief efforts. In larger communities, state and federal agencies GIS operations have become an integral part of Emergency Operation Centers (EOCs), but in some instances, e.g. smaller and/or rural communities, special GIS operators might not be available or are not part of the Emergency Operation Center staff. Even in New Orleans, a major metropolitan area, GIS use was hindered during Hurricane Katrina because the mapping requests overwhelmed the EOC capabilities and outdated computers caused frustration [1].

## II. EVACUATION PLANNING

Crises mean the loss of enterprises and individuals, and therefore losses in the sum total of community development, crises have several types such as nature, industrial, technological, etc. Crisis management is the process to control the crisis by developing plans to reduce the risk of a crisis occurring and to deal with any crises that arise, and the implementation of these plans so as to minimize the impact of crises and assist the organization to recover from them.

Evacuation is one of the crises management activities which is an operation where by all or part of a particular population is temporary relocated, whether spontaneously or in an organized manner, from a sector that has been struck by a disaster or is about to be struck by a disaster, to a place considered not dangerous for its health or safety [11]. Evacuation can be carried for several reasons such as volcanoes, floods, hurricanes, earthquakes, military attacks, industrial accidents, traffic accidents, fire, nuclear accidents … etc.

Evacuation Plan is a supporting document that is used to identify and organize the various responses aimed at evacuating persons exposed to a threat from an evacuation sector to a reception sector, while ensuring them a minimum of essential services during an emergency. Proper planning will use multiple exits, contra-flow lanes, and special technologies to ensure full, fast and complete evacuation and should consider personal situations which may affect an individual's ability to evacuate. These plans may also include alarm signals that use both aural and visual alerts.

All countries should have a written evacuation plan in order to facilitate a safe and efficient evacuation or relocation for their people and the plan must be updated regularly. Country Directors must communicate in writing what evacuation assistance will provide for each member of the staff and their families in the event of a crisis [2].

Evacuation planning has been studied from different perspectives such as evacuee's behaviors, traffic control, safe area selection, and route finding to safe areas [3], with any perspective an evacuation plan should involve four phases [2]:

1. **Pre-Planning:** During this phase, operations are normal with periodic update. The country office must ensure continual monitoring of the safety and security situation, especially in high risk areas.

2. **Alert:** Mounting tension may lead the country director to issue a recommendation to limit operations, increase security measure, and review the evacuation plan.

3. **Curtailment of operations/relocation:** The situation has deteriorated to a level unsafe for normal operations and may require rapid evacuation.

4. **Evacuation:** The planned evacuation process become in effect and all threatened people must be transferred to safe areas.

### III.  GEOGRAPHICAL INFORMATION SYSTEMS

As sun rises and sets, people everywhere ask questions about locations on earth like: Where can I find the shop? Where is the nearest library? How can I go to the restaurant? Which site is the best site for the building? All of these questions and more can be asked by a Geographical Information System (GIS). GIS involves handling the issues arising from working with geographic information, also it examines the effect of GIS on people and society, and the effect of society on GIS.

There have been so many attempts to define GIS that make it difficult to select one definitive definition [4] because the definition will depend on the one giving it and his point of view. For example, *Rhind D.W. (1989)* defines it generally as "a computer system that can hold and use data describing places on the Earth's surface" [5]. Other definitions explain what a GIS can do. For example, *Burrough P.A. (1986)* define a GIS as "a set of tools for collecting, storing, retrieving at will, transforming, and displaying spatial data from the real world for a particular set of purposes" [6], the US Government defines it as "a system of computer software and procedures designed to support the capture, management, manipulation, analysis, and display of spatially referenced data for solving complex planning and management problems." [12], and the *Department of the Environment (1987)* say that a GIS is "a system for capturing, storing, checking, integrating, manipulating, analyzing and displaying data which are spatially referenced to the earth" [7].

Simply, GIS system considers three main components: hardware, software, and spatially referenced data. In particular, a working GIS needs to integrate five components:

*1.  Hardware*

GIS hardware includes a computer with high capabilities on which a GIS operates, a monitor on which results displays, a printer to display results as reports, other GIS hardware also includes GPS instrument to collect coordinates, and a digitizer.

*2.  Software*

Key software components are:

- System software (e.g. operating system).
- a database management system (DBMS)
- tools for the input and manipulation of geographic information
- tools that support geographic query, analysis, and visualization
- a graphical user interface (GUI) for easy access to tools
- Drawing software.

*3.  People*

GIS people can be divided into two main categories:

- People who develop the GIS and define its tasks such as database administrators, application specialists, systems analysts, and programmers. They are responsible for maintenance of the geographic database and provide technical support.
- General users who are using the GIS in their daily business.

*4.  Data*

May be the most important part of a GIS, a GIS integrates spatial data with other attribute data to answer unique spatial queries provided by users.

- Spatial data: data can be referenced to a location on earth. For example, country, road, river, etc.
- Attribute data: also called aspatial data, data linked to spatial data describe those data. For example, country name, road length, river width, etc.

*5.  Methods*

A well defined consistent rules that the GIS needs to achieve its goals includes how the data will be retrieved, input into the system, stored, managed, transformed, analyzed, and finally presented in a final output.

From information systems point of view, GIS is like any other information system that considers three main phases: input, processing, and output. GIS helps in answering questions and solving problems by looking at the data in a way that is quickly understood and easily shared. GIS technology can be integrated into any enterprise information system framework [13].

What distinguish GIS system from any other information system is that it combines a powerful visualization environment with a strong analytical and modeling framework, which makes GIS attractive to most people in the whole world. For example, when a rainfall occur it is important to know where it is located. By using a spatial reference system such as latitude, longitude, or elevation we can know where the rainfall, and by comparing the results with the landscape one can predicates if there are someplace that likely to be subjected to dry up.

Designing a GIS system is a sequenced process begins with the data collection phase in which data; spatial and attribute data, from various data sources is collected, these spatial data must be geo-referenced to their spatial location on the earth. Then these data are digitized to convert it from the analog format into the digital one by a trace method. After that, attribute data is combined with the spatial data into a data format so that it can be manipulated by the system to provide answers that help decision makers take their decisions. This process is utilized in Figure 1.

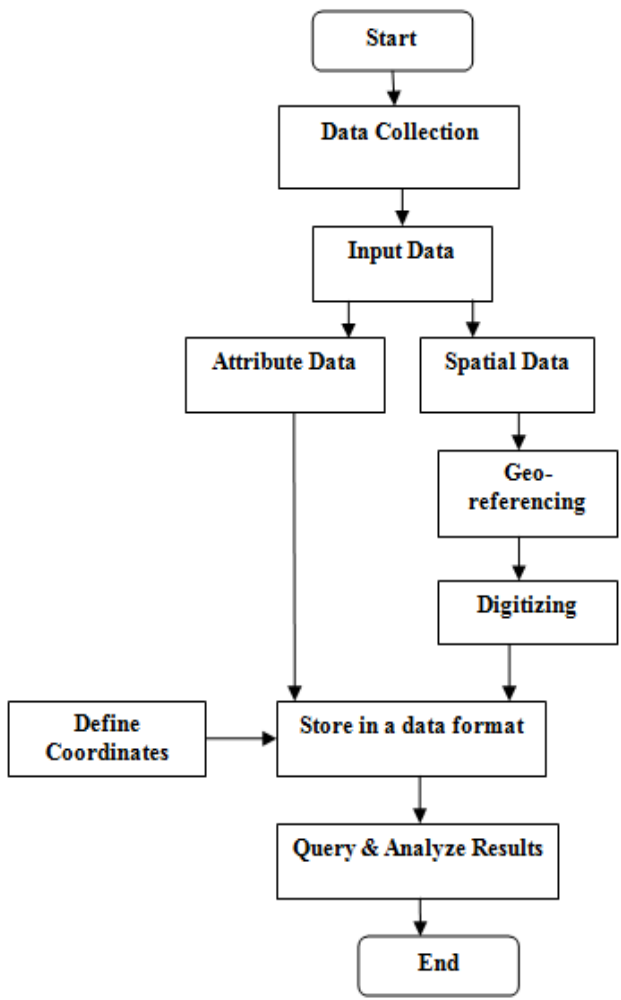

Figure 1. GIS system's design process.

## IV. USING GIS IN EVACUATION PLANNING

Evacuation is a process in which threatened people are transferred from the incident place to a safer place to protect their lives. It is a very complex process, besides needing to be accurate and carful; it must be done very quickly. GIS plays an important role in emergency management in general and in evacuation planning in particular.

In 2001 GIS was prominently used in the rescue, relief and recovery process after the World Trade Center attack. Although New York City's Emergency Operation Center and GIS infrastructure was destroyed, city officials were able to set up a backup facility and use GIS to produce maps for emergency response purposes by the evening of 9/11[1]

*Yang Bo et al (2009)* assure that emergency evacuation is an important measure for preventing and reducing injuries and death during large scale emergency. They assumed that the efficiency of evacuation is based on (1) Understanding of the situation, and (2) Good analysis and judgment of information.

So they proposed a multi-agent framework and a GIS system that:

1. Simulate individual movement by a modified Particle Swarm Optimization (LWDPSO) which considers each individual as a particle and as one particle found an exit all other particles should consider all other exits as an exit and choose the nearest one, and

2. Modeling the evacuation environment by a GIS platform, in which each individual takes an average space 0.4m × 0.4m when it is very crowded, and then build a potential map for evacuation environment which describes the distance between an individual and an exit [8].

*Michel Pidd et al (1993)* developed a Configurable Evacuation Management and Planning System (CEMPS) to be used for evacuation from man-made disasters. They found GIS as an efficiency technology that can examine static aspects of an evacuation plan such as determining evacuation zone and evacuation routes, but they assume that it can't consider dynamic aspects such as How long vehicles take to come? How long will it take to evacuate the population? So they use micro-simulation method in order to simulate the movement of people from the evacuation zone to the vehicles and use ArcInfo to determine the evacuation zone and evacuation routes [9].

*Mohammad Saadatseresht et al (2008)* introduce that the distribution of population into safe areas during evacuation process is a vital problem that affect the efficiency of the evacuation plan. They propose a three step approach in order to determine the distribution of evacuees into the safe areas, in which step1: is to select the safe areas, step2: is to find optimal path between each building block and the candidates safe areas, and step3: is to select optimal safe areas for each building block, optimal safe area should be the closest to the building block and should have enough space for evacuees. To achieve the third step two objective functions were defined, and then the spatial MOP was solved using the NSGA-II algorithm in a GIS environment [3].

*Bo Huang and Xiaohong Pan (2006)* introduce an Incident Response Management Tool (IRMT) in order to reduce response time in incident management. The IRMT consists of:
1. GIS system: provide user interface, process network data, find shortest path, and visualize the result.
2. Traffic simulation engine: simulate incidents, gather link travel time at regular intervals, and send this time to the GIS system.

Optimization engine: to minimize the overall travel time of all response units [10].

## ACKNOWLEDGMENT

First of all, Praise and thank Allah that his grace is righteous, and then I thank my family and my friends for supporting me.